\documentstyle[agupp,psfig]{article} 
\lefthead{HELMSTETTER and SORNETTE}
\righthead{Foreshock as Triggered Seismicity}	
        \authoraddr{ Agn\`es Helmstetter, Institute of
Geophysics and Planetary Physics, University of California, Los
Angeles (e-mail: helmstet@moho.ess.ucla.edu)}
\authoraddr{Didier Sornette,
Department of Earth and Space Sciences and Institute of
Geophysics and Planetary Physics, University of California, Los Angeles,
California and Laboratoire de Physique de la Mati\`{e}re Condens\'{e}e,
CNRS UMR 6622 and Universit\'{e} de Nice-Sophia Antipolis, Parc Valrose,
06108 Nice, France (e-mail: sornette@moho.ess.ucla.edu)}

\input{update.tex}
\begin{document}
\title{Foreshocks Explained by Cascades of Triggered Seismicity}
\author{Agn\`es Helmstetter}
\affil{ Institute of
Geophysics and Planetary Physics, University of California, Los Angeles,
       California 90095-1567.}
\author{Didier Sornette}
\affil{Department of Earth and Space Sciences and Institute of
Geophysics and Planetary Physics, University of California, Los Angeles,
       California 90095-1567 and
     Laboratoire de Physique de la Mati\`{e}re Condens\'{e}e, CNRS UMR 6622
Universit\'{e} de Nice-Sophia Antipolis, Parc Valrose, 06108 Nice, France
}

\newcommand{\be}{\begin{equation}}
\newcommand{\ee}{\end{equation}}
\newcommand{\ba}{\begin{eqnarray}}
\newcommand{\ea}{\end{eqnarray}}
\newenvironment{technical}{\begin{quotation}\small}{\end{quotation}}
\renewcommand{\thefootnote}{\arabic{footnote}}
\newtheorem{definition}{Hypothesis}


\begin{abstract}

The observation of foreshocks preceding large earthquakes and the
suggestion that foreshocks have specific properties that may
be used to distinguish them from other earthquakes have raised the hope
that large earthquakes may be predictable. Among proposed anomalous properties
are the larger proportion than normal of large versus small foreshocks,
the power law acceleration of seismicity rate as a function of
time to the mainshock and the spatial migration of foreshocks toward
the mainshock, when averaging over many sequences.
Using Southern California seismicity, we show that these properties and others
arise naturally from the simple model that any earthquake may trigger
other earthquakes, without arbitrary distinction between foreshocks,
aftershocks and mainshocks.
We find that foreshocks precursory properties are independent of the
mainshock size. This implies that earthquakes (large or small) are
predictable to the same degree as seismicity rate is predictable
from past seismicity by taking into account cascades of triggering.
The cascades of triggering give rise naturally to long-range
and long-time interactions, which can explain the observations
of correlations in seismicity over surprisingly large length scales.

\end{abstract}

\begin{article}

\section{Foreshocks, mainshocks and aftershocks: hypothesis and predictions}

It has been recognized for a long time that large earthquakes are
sometimes preceded by an acceleration of the seismic activity, known
as foreshocks \cite{JM76,AM96}.
In addition to the increase of the seismicity rate a few hours to months before
large earthquakes, other properties of foreshocks have been reported, which
suggest their usefulness (when present) as precursory patterns for earthquake
prediction.
Special physical mechanisms have been proposed for foreshocks with the hope
of helping earthquake prediction hope of helping earthquake prediction
\cite{Yama2,sorvakn,Dodgeg,DK96}.
In addition, anomalous precursory seismic activity
extending years to decades before large earthquakes and at distances
up to ten times the mainshock rupture size are often thought
to require different physical mechanisms
\cite{KeiMal,KnopoffetalKei,Bowmanetal,JS,Sorsammis,KBstate}
than for foreshocks closer to the mainshock epicenters.

The division between foreshocks, mainshocks, and
aftershocks has a long and distinguished history in seismology.
Within a pre-specified space-time domain, foreshocks are usually defined as
earthquakes (above the background rate) preceding a larger earthquake
(mainshock), which is itself followed by an increase in seismicity of smaller
earthquakes (aftershocks).
However, recent empirical and theoretical scrutiny suggests that
this division might be arbitrary and physically artificial
\cite{KK81,Shaw,Jones95,Houghjones,Felzer}.
Since the underlying physical processes are not fully understood,
the qualifying time and space windows used to select aftershocks, mainshocks
and aftershocks are more based on common sense than on hard science.
If the space-time window is extended and
a new event not considered previously is found
with a magnitude larger than the previously classified mainshock,
it becomes the new mainshock and all preceding events are retrospectively
called foreshocks.
A clear identification of foreshocks, aftershocks and mainshocks is
hindered by the fact that nothing distinguishes them in their seismic
signatures: at the present level of resolution of seismic inversions,
they are found
to have the same double-couple structure and the same radiation patterns
\cite{Houghjones}. Statistically, the aftershock magnitudes are
distributed according
to the Gutenberg-Richter (GR) distribution $P(m) \sim 10^{-bm}$
with a $b$-value similar to other earthquakes \cite{ranalli,Utsu}.
However, some studies \cite{KKK82,Molchan90,Molchan99} have suggested that
foreshocks have a smaller $b$-value than other earthquakes, but the physical
mechanisms are not yet understood.

\begin{figure}
\begin{center}
\psfig{file=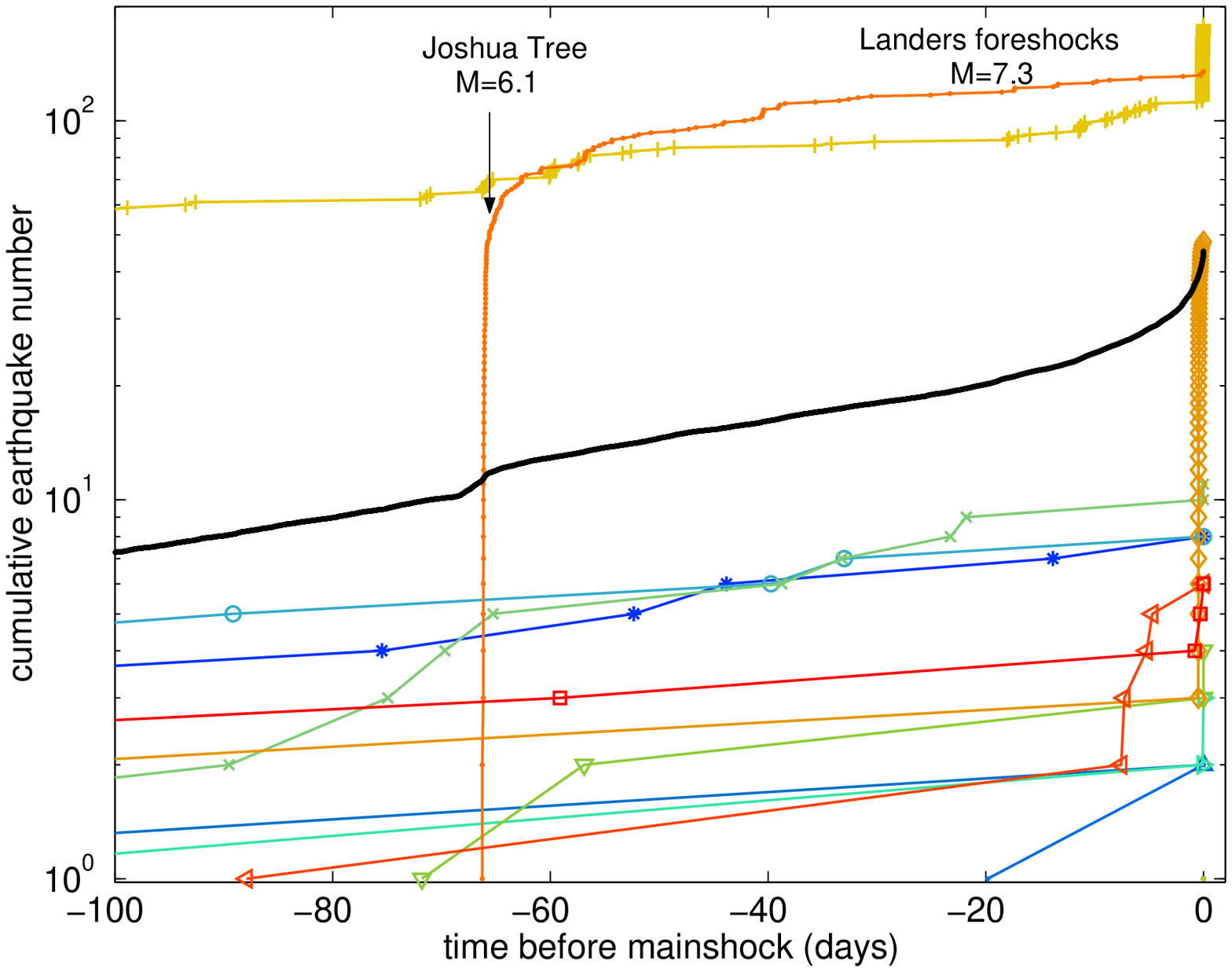,width=14cm}
\caption{\label{fortyp2} Cumulative number of foreshocks of type II
(see definition in section \ref{techfore}) for all $M\geq 6.5$
mainshocks (thin gray lines), and average number of foreshocks per
mainshock (heavy black line), obtained by stacking all (3700)
foreshock sequences of $M\geq 4$ mainshocks in the SCEC catalog.
The foreshocks have been
selected in a time-space window with $T=200$ days and $R=30$ km.
}
\end{center}
\end{figure}

The Omori law describes the power law decay of the aftershock rate
$\sim 1/(t-t_c)^{p}$ with time from a mainshock that occurred at $t_c$
\cite{Omori,KK78,Utsu}, and which may last from months up to decades.
In contrast with the well-defined Omori law for aftershocks,
there are huge fluctuations of the foreshock seismicity rate, if any, from one
sequence of earthquakes to another one.
Figure \ref{fortyp2} shows foreshock sequences of all $M\geq6.5$
mainshocks in the catalog of Southern California seismicity
provided by the Southern California Seismic Network for the period 1932-2000.
By stacking many foreshock sequences, a well-defined acceleration of
the seismicity preceding mainshocks emerges,
quantified by the so-called inverse Omori law $\sim 1/(t_c-t)^{p'}$,
where $t_c$ is the time of the mainshock \cite{Papazachos,KK78,JM79}.
While we see clearly an acceleration for the averaged foreshock number
  (black line in Figure  \ref{fortyp2}), there are huge fluctuations
of the rate of foreshocks for individual sequences.
Most foreshock sequences are characterized by the occurrence of a major
earthquake before the mainshock, which has triggered the mainshock
most probably indirectly due to a cascade of multiple triggering.
For instance, 66 days before its occurrence, the $M=7.3$ Landers
earthquake (upper curve  in Figure  \ref{fortyp2}) was preceded by the
  $M=6.1$ Joshua-Tree earthquake. The successive oscillations of the
cumulative number of events after the Joshua-Tree earthquake correspond
to secondary, tertiary, etc., bursts of triggered seismicity.
The inverse Omori law is usually observed for time scales
shorter than the direct Omori law, of the order up to weeks to
a few months before the mainshock.
However,  there seems to be no way of identifying foreshocks
from usual aftershocks and mainshocks in real time
(see \cite{Jones95,Houghjones} for a pioneering presentation of this
view point). In other words, is the division between foreshocks,
mainshocks and aftershocks falsifiable \cite{Kaganfalsi}?

In order to address this question, we present a novel
analysis of seismic catalogs, based on a parsimonious
model of the foreshocks, mainshocks and aftershocks,
in terms of earthquake triggering: earthquakes may trigger
other earthquakes through a variety of physical mechanisms \cite{Harris}
but this does not allow one to put a tag on them.
Thus, rather than keeping the specific classification that foreshocks
are precursors of mainshocks and mainshocks
trigger aftershocks, we start from the following hypothesis:

\paragraph*{Hypothesis}
\label{hy}
{\it Foreshocks, mainshocks, and aftershocks are
physically indistinguishable.}

In this paper, we study a simple model of seismicity based on this Hypothesis
and demonstrate that it can explain many properties of foreshocks, such as
the larger proportion than normal of large versus small foreshocks,
the power law acceleration of seismicity rate as a function of
time to the mainshock and the spatial migration of foreshocks toward
the mainshock, when averaging over many sequences. These properties and
others arise naturally from the simple model that any earthquake may trigger
other earthquakes, without arbitrary distinction between foreshocks,
aftershocks and mainshocks.

\subsection{Definition of the model}
\label{ETAS}
The simplest construction that embodies the Hypothesis is
the epidemic-type  aftershock sequence (ETAS) model introduced
in \cite{KK81,KK87} (in a slightly different form) and in \cite{Ogata88}.
In this model, all earthquakes may be simultaneously
mainshocks, aftershocks and possibly foreshocks. An observed ``aftershock''
sequence in the ETAS model is the sum of a cascade of events in which
each event can trigger more events. The triggering process
may be caused by various mechanisms
that either compete or combine, such as pore-pressure changes due
to pore-fluid flows coupled with stress variations, slow redistribution of
stress by aseismic creep, rate-and-state dependent friction within faults,
coupling between the viscoelastic lower crust and the brittle upper crust,
stress-assisted micro-crack corrosion, etc..

The epidemic-type aftershock (ETAS) model assumes that a given event
(the ``mother'') of magnitude $m_i$ occurring at time $t_i$ and position
$\vec r_i$ gives birth to other events (``daughters'') of any possible
magnitude $m$ at a later time between $t$ and $t+dt$ and at point
$\vec r \pm \vec dr$ at the rate
\be
\phi_{m_i}(t-t_i, \vec r-\vec r_i) = \rho(m_i)~\Psi(t-t_i)~\Phi(\vec
r-\vec r_i)~.
\label{first}
\ee
We will refer to $\phi_{m_i}(t-t_i, \vec r-\vec r_i)$ as the ``local'''
Omori law, giving the seismic rate induced by a single mother.
It is the product of three independent contributions:
\begin{enumerate}
\item $\rho(m_i)$ gives the number of daughters born from a mother with
magnitude $m_i$. This term is in general chosen to account for the
fact that large earthquakes have many more triggered events than small
earthquakes.  Specifically,
\be
\rho(m_i) = K ~10^{\alpha m_i}~,  \label{formrho}
\ee
which
  is in good agreement with the observations of aftershocks productivity
(see {\it Helmstetter} [2003] and references therein).

\item $\Psi(t-t_i)$ is a normalized waiting time distribution giving the rate
of daughters born at time $t-t_i$ after the mother
\be
\Psi(t) = {\theta~c^{\theta}  \over (t+c)^{1+\theta}}~.
\label{psidef}
\ee
The ``local'' Omori law  $\Psi(t)$ is generally significantly different from
the observed Omori law of aftershocks.
We have shown in [{\it Helmstetter and Sornette}, 2002a] that the exponent
$p$ of the observed ``global'' Omori law is equal to or smaller than
the exponent $1+\theta$ of the ``local'' Omori law  $\Psi(t)$  due to the
cascades of triggering.

\item $\Phi(\vec r-\vec r_i)$ is a normalized spatial ``jump'' distribution
from the mother to each of her daughter, quantifying the probability
for a daughter
to be triggered at a distance $|\vec r-\vec r_i|$ from the mother.
Specifically, we take
\be
\Phi(\vec r) = {\mu \over d ({|\vec r| \over d}+1)^{1+\mu}}~,
\label{phidef}
\ee
where $\mu$ is a positive exponent and  $d$  is a characteristic dimension
of the earthquake source, as suggested by [{\it Ogata}, 1999].
\end{enumerate}

The last ingredient of the ETAS model is that the magnitude $m$ of
each daughter is chosen independently from that of the mother and of
all other daughters according to the Gutenberg-Richter distribution
\be
P(m) = b~ \ln(10) ~ 10^{-b (m-m_0)}~, \label{gojfwo}
\ee
with a $b$-value usually close to $1$.
$m_0$ is a lower bound magnitude below which no daughter is triggered.

The total seismicity rate is the sum of all aftershocks
of past events and of a constant external source $s$ used to model
the constant tectonic loading.
The full distribution of seismicity in time, space and magnitude is
thus described by the 8 parameters of the ETAS model $K$, $c$, $\theta$,
$\mu$, $d$, $\alpha$, $b$ and $s$. See [{\it Ogata}, 1999],
[{\it Helmstetter and Sornette}, 2002a] and references therein for a
discussion of the values of these parameters in real seismicity.

The ETAS model has been used previously to give short-term probabilistic
forecast of seismic activity \cite{KK87,KJ,CM01}, and to describe the
temporal and spatial clustering of seismic activity
\cite{Ogata88,K91,KJ,CM01,Felzer}. The Hypothesis and the ETAS model
allow us to study two classes of foreshocks, called of type I
and of type II defined in the next section.

\subsection{Definition of foreshocks of type I and of type II }
  \label{techfore}
\subsubsection{Formal definitions}
The usual definition of foreshocks, that we shall call
``foreshock of type I,'' refers to any event of magnitude smaller
than or equal to the magnitude of the following event, then identified
as a ``mainshock.'' This definition implies the choice of a space-time
window $R \times T$ used to define both foreshocks and mainshocks.
Mainshocks are large earthquakes that were not preceded by a larger event
in this space-time window.

In contrast, the Hypothesis that the same physical mechanisms describe
the triggering of  a large earthquake  by a smaller earthquake
(mainshock  triggered by a type I foreshock) and the triggering
of a small earthquake by a larger event (aftershock triggered by a
mainshock) makes it natural to remove the constraint that foreshocks must
be smaller than the mainshock. We thus define  ``foreshock  of type II''
as any earthquake preceding a large earthquake which is defined as the
mainshock, independently of the relative magnitude of the foreshock
compared to that of the mainshock. That is, foreshocks of type II may
actually be larger than their mainshock.

The advantage of this second definition is that foreshocks of type II are
automatically defined as soon as one has identified the mainshocks,
for instance, by calling mainshocks all events of magnitudes larger than
some threshold of interest.
Foreshocks of type II are thus all events preceding these large
magnitude mainshocks in a space-time window  $R \times T$.
In contrast, foreshocks of type I need to obey a
constraint on their magnitude, which may be artificial, as we shall see
further down.

\subsubsection{Practical implementation}

In our analysis of the SCEC catalog, we construct foreshock and aftershock
sequences as follows. A mainshock is defined as an earthquake in the magnitude
range $(M,M+\Delta M)$ that was not preceded by
a larger event in a space-time window ($R_2$, $T$) before the mainshock.
The distance $R_2 \cong 50$ km is here chosen to be close to (but
smaller than) the maximum size of the spatial clusters
of seismicity in the California catalog, in order to minimize the influence
of large earthquakes that occurred before the mainshock. Other choices
between 20 km to 200 km have been tested and give essentially the same results.
The aftershocks are all events that occurred in a space-time window
($R$, $T$) after each  mainshock.
The foreshocks of type I are selected in a space-time window  $R$, $T$ before
each mainshock.
  Foreshocks of type II are selected in the same space-time window ($R$, $T$)
before each mainshock, now defined
without the constraint that they were not preceded by
a larger event in a space-time window ($R_2$, $T$). The only difference
between foreshocks of type I and of type II is that the selection of
their respective mainshocks is different.

The minimum magnitude cutoff used for aftershocks and foreshocks is $M_0=3$.
The fact that the SCEC catalog is not complete below $M=3.5$ for the first
part of the catalog 1932-1975 does not affect the results on the
inverse and direct Omori laws, it simply under-estimates the
number of foreshocks and aftershocks but does not change the temporal
evolution of the rate of seismicity before and after mainshocks.
The incompleteness of the catalog for $M<3.5$  before 1975 explains the
roll-off of the magnitude distribution for small magnitudes.
Even for the most recent part, the SCEC catalog is not
complete above $M=3$ in the few hours or days after a large
$M\geq6$ earthquake, due to the saturation of the seismic network.
These missing events may induce a spurious roll-off of both the direct and
inverse Omori laws at short times before or after a mainshock, but cannot
induce a spurious acceleration of the rate of seismicity before mainshocks.

We stack all foreshocks and aftershocks sequences synchronized at the time
of the ``mainshocks'' in different mainshock magnitude intervals
[$M$,$M+\Delta M$], and for different choices of the space-time window
$R$, $T$ used to define foreshocks and aftershocks. We use larger
magnitude intervals $\Delta M$ for larger mainshock magnitudes $M$ to
compensate for the smaller number
of large mainshocks. $R$ has been tested between $10$ km and
up to $500$ km with no essential change, except for an increasing
sensitivity to the background seismicity for the largest $R$ (see
Figure \ref{forncR}).
$T$ has been tested between $0.5$ year to $10$ years with similar results.
Tests have also been performed with the spatial window size $R$
adjusted to scale with the mainshock magnitude with no significant difference.
Our results presented below thus appear robust with respect
to the (arbitrary) definitions of the space-time windows and the definition
of mainshocks. In the main text, foreshocks of type I
and of type II are treated separately.
Previous studies of foreshocks using a stacking method
\cite{Papa75a,Papa75b,JM76,KK78,JM79,Shaw,Reas99}
have considered foreshocks of type I only.

\subsection{Foreshock properties derived from the ETAS model}

The simple embodiment of the Hypothesis in the ETAS model leads
to the following consequences and predictions \cite{Helmfore}, which
are proposed as crucial tests of the Hypothesis.
\begin{enumerate}
\item The rate of foreshocks of type II is predicted to increase before the
mainshock according to the inverse Omori law $N(t) \sim 1/(t_c-t)^{p'}$
with an exponent $p'$ smaller than the exponent $p$ of the direct Omori law.
The exponent $p'$ depends on the ``local'' Omori exponent $1+\theta$
describing the direct triggering rate between earthquakes
(first-generation triggering),
on the $b$-value of the GR distribution and on the exponent $\alpha$
quantifying the increase $\propto 10^{\alpha M}$ in the number of aftershocks
as a function of the magnitude $M$ of the mainshock \cite{alpha}. The
inverse Omori law
also holds for foreshocks of type I preceding large mainshocks.
The inverse Omori law results from the direct Omori law for
aftershock and from the multiple cascades of triggering.
The inverse Omori law emerges as the
expected (in a statistical sense) trajectory of seismicity, conditioned
on the fact that it leads to the burst of seismic activity
accompanying the mainshock.

\item In contrast with the direct Omori law, which is clearly observed after
all large earthquakes, the inverse Omori law is observed only when stacking
many foreshock sequences. Even for small mainshocks, for which the number
of aftershocks is similar to the number of foreshocks, there are much larger
fluctuations in the rate of foreshocks before mainshocks than in
the rate of aftershocks.

\item While the number of aftershocks increases as $10^{\alpha M}$
with the magnitude $M$ of the mainshock,
the number of foreshocks of type II is predicted to be independent of $M$.
Thus, the seismicity should increase on average
according to the inverse Omori law before any earthquake, whatever its
magnitude. For foreshocks of type I, the same results should hold for large
mainshocks. For small and
intermediate values of the mainshock magnitude $M$,
the conditioning on foreshocks of type I to be smaller
than their mainshock makes their number increase with $M$ solely due to the
constraining effect of their definition.

\item The GR distribution for foreshocks is predicted
to change upon the approach of the mainshock, by
developing a bump in its tail. Specifically, the modification of the
GR law is predicted to take the shape of an additive
correction to the standard power law, in which the new term is
another power law with exponent $b-\alpha$.
The amplitude of this additive power law term is predicted
to also exhibit a power law acceleration upon the approach to the mainshock.

\item The spatial distribution of foreshocks is predicted to
migrate toward the mainshock as the time increases toward the time
of the mainshock, by the mechanism of a cascade of seismic triggering
leading to a succession of jumps like in a continuous-time random
walk \cite{etasdif}.

\end{enumerate}

We now proceed to test systematically these predictions
on the catalog of the Southern California Data Center (SCEC)
over the period 1932-2000, which is almost complete above $M=3$
and contains more than 22000 $M\geq3$ earthquakes, using the
methodology described in section \ref{techfore}.

\section{Properties of foreshocks, mainshocks and aftershocks and
   comparison with the ETAS model}

\subsection{Observation of the direct and inverse Omori law}
  \label{obs:ominv}
\begin{figure}
\begin{center}
\psfig{file=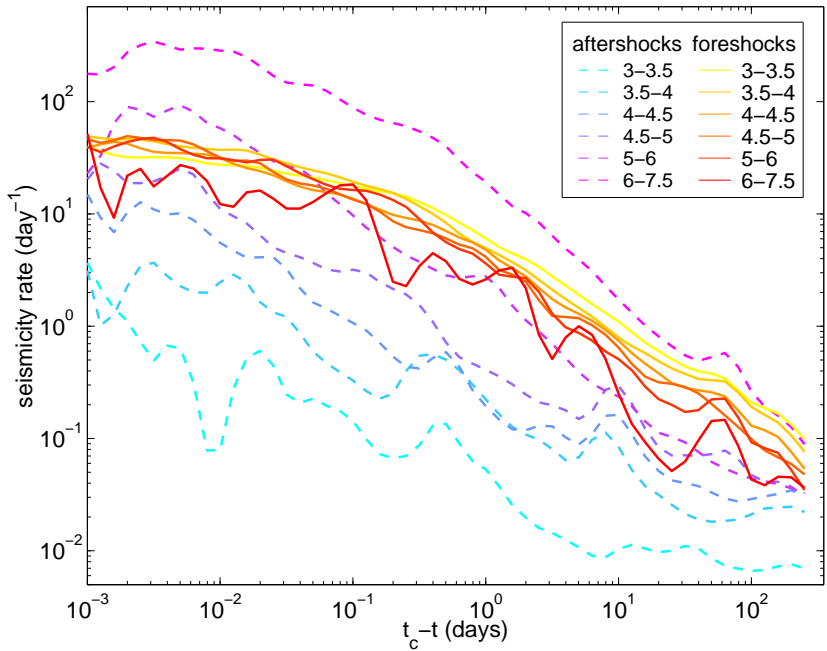,width=11cm}
\caption{\label{SCECAFNCR50T}
Rate of seismic activity per mainshock for foreshocks of type II
(continuous lines) defined in section \ref{techfore}
and for aftershocks (dashed lines) measured as a function of the time
$|t-t_c|$ from the mainshock occurring at $t_c$, obtained
by stacking many earthquake sequences for different mainshock magnitude
intervals given in the legend.
This figure illustrates the power-law increase of seismicity before
the mainshock (inverse Omori law) and the power-law decrease
after the mainshock (direct Omori law), and show how the rate
of foreshocks and aftershocks changes with mainshock magnitude.
}
\end{center}
\end{figure}

Figure \ref{SCECAFNCR50T} shows the rate of foreshocks of type II as
a function of $t_c-t$ and of aftershocks as a function of $t-t_c$, where
$t_c$ is the time of the mainshocks, for different mainshock magnitude
between 3 and 7. Aftershocks and foreshocks have been selected
with the space-time window $T=1$ yr and $R=50$ km.
Both rates follow an approximate power law
(inverse Omori law for foreshocks with exponent $p'$ and Omori law
for aftershocks with exponent $p$).

The fluctuations of the rate of foreshocks are larger for large mainshocks,
because the number of mainshocks (resp. foreshocks) decreases from $15584$
(resp.  $1656249$) for the magnitude range $3-3.5$ down to $47$ (resp. $1899$)
for $M>6$ mainshocks. In contrast, the fluctuation of the rate of aftershocks
are larger for small mainshocks magnitudes, due to the increase of the number
of aftershocks per mainshock with the mainshock magnitude, and to the rules
of mainshock selection which reject a large proportion of small earthquakes.
The number of mainshocks is different for aftershocks
and foreshocks due to their distinct definition.
The number of mainshocks associated with aftershocks decreases from $677$
for the magnitude range $3-3.5$ down to $39$ for $M \geq 6$ mainshocks.
The number of aftershocks increases from $2614$ for $3 \leq M <3.5$ up to
$7797$ for $M \geq6$ mainshocks. The fluctuations of the data
makes it hard to exclude the hypothesis that the two power
laws have the same exponent $p =p' \cong 1.$, even if $p'$ seems
slightly smaller. Note that it can be shown theoretically for $\alpha < b/2$
that $p=1-\theta$ and $p'=1-2 \theta$ for a local Omori law with exponent
$1+\theta$ and that the difference $p-p'$ should get smaller as $\alpha$
increases above $b/2$ \cite{Helmfore}. Since $\alpha \cong 0.8 > b/2
\approx 0.5$,
this limit is met which explains the smallness of the difference $p-p'$.
The truncation of the seismicity rate for small times $|t_c-t| < 1$ day,
especially for aftershocks of large $M>6$ mainshocks and for foreshocks,
is due to the incompleteness of the catalog at very short times after
mainshocks due to the saturation of the seismic network.
At large times from the mainshock, the seismicity rate decreases to
the level of the background seismicity, as seen clearly for the
rate of aftershocks following small $M=3$ mainshocks.

The second striking observation is the strong variation of the amplitude of the
rate $N_a(t)$ of aftershocks as a function of the magnitude $M$ of
the mainshock, which is well-captured by an exponential dependence
$N_a(t) \propto 10^{\alpha M}/(t-t_c)^p$ with $\alpha \cong 0.8$ \cite{alpha}.
In contrast, the rates of foreshocks of type II are completely
independent of the magnitude $M$ of the mainshocks: quite strikingly,
all mainshocks
{\it independently} of their magnitudes are preceded by the same statistical
inverse Omori law, with the same power law increase and the same
absolute amplitude!
All these results are very well modeled by the ETAS model
with the parameters $\alpha=0.8$, $\theta=0.2$ and $b=1$ using
the theoretical framework and numerical simulations developed in
\cite{Helmfore}.

\begin{figure}
\begin{center}
\psfig{file=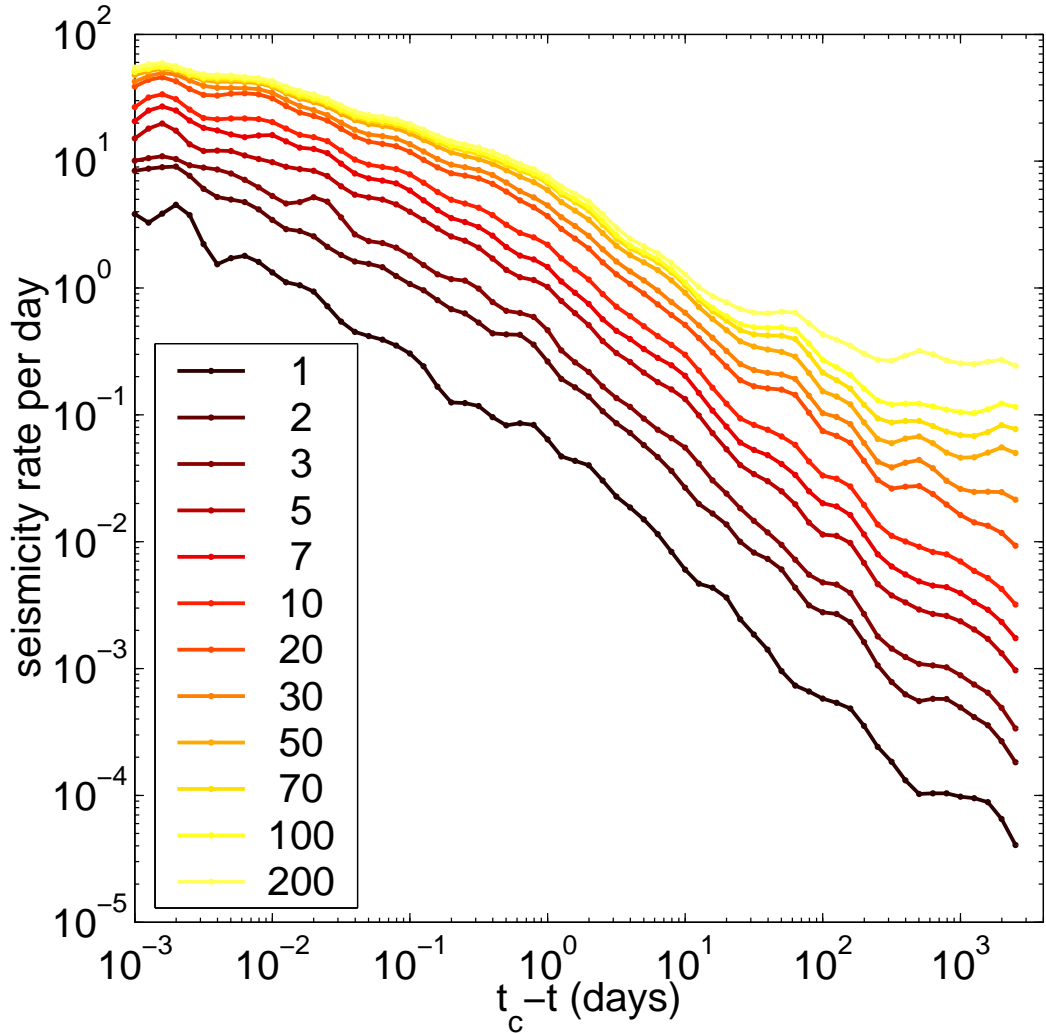,width=14cm}
\caption{ \label{forncR} Rate of foreshocks of type II (defined in
  section \ref{techfore}) averaged over 2158  mainshock with magnitudes
in the range $(4,4.5)$, for $T=10$ yrs and for different choices of the
distance $R$ between 1 and 200 km used to select foreshocks around mainshocks.
The total number of foreshocks of type II increases from 1001 for
$R=1$ km up to 2280165  for $R=200$ km.}
\end{center}
\end{figure}

\begin{figure}
\begin{center}
\psfig{file=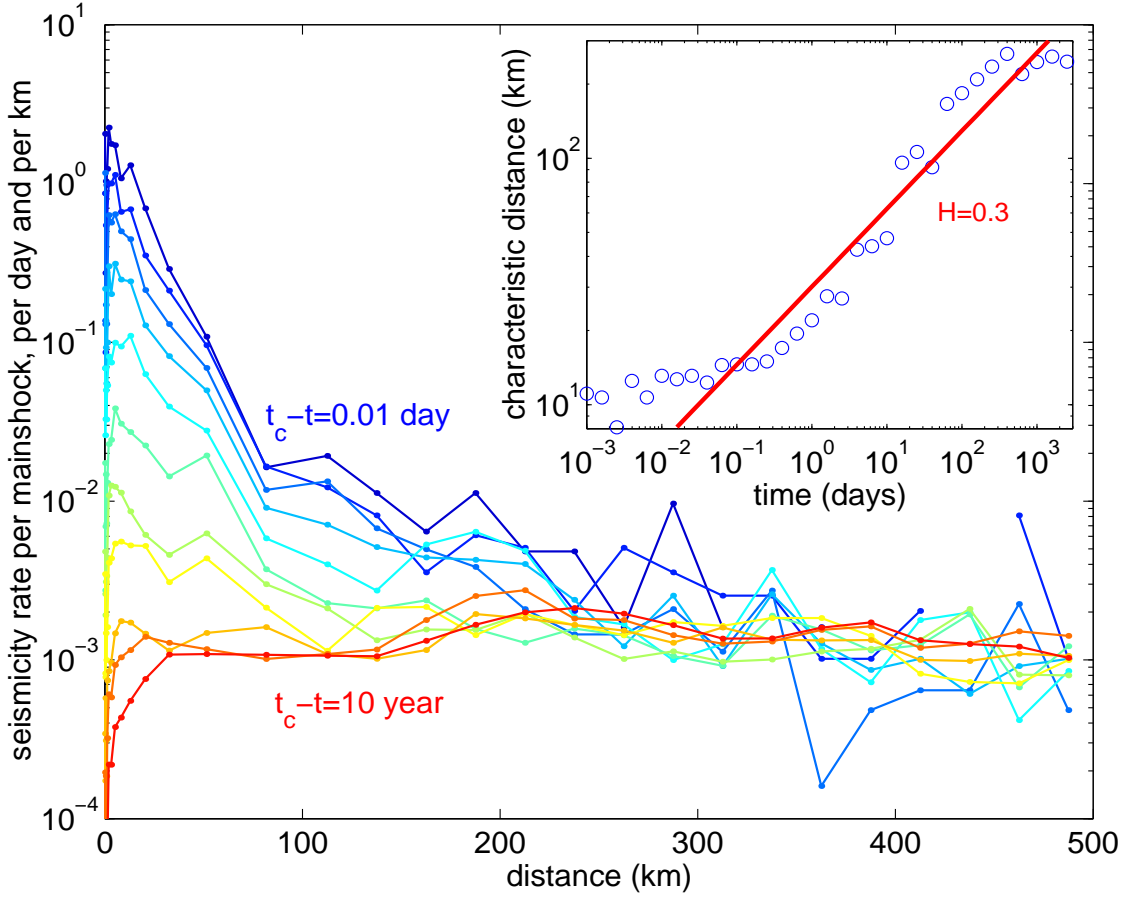,width=15cm}
\caption{\label{Prfor}
Rate of foreshocks of type II (defined in section \ref{techfore})
before $M\geq4.5$ mainshocks as a function of the distance from
the  mainshock for different values of the time before the mainshock
ranging from 0.01 day (black line at the top) to 10 yrs
(gray line at the bottom).
We use logarithmic bins for the time windows, with a bin size
increasing from 0.01 day up to 10 yrs as a geometric series with
multiplicative factor 3.2.
The number of events in each time window increases from  $N=936$ for
$0.01- 0.03$ days up to $2096633$ for $1000-3650$ days.
We evaluate the seismicity rate for different distances from the mainshock by
counting the number of events in each shell $(r,r+\Delta r)$.
The seismicity rate is normalized by the number of mainshocks, the
duration of the time window and the widths of the space window $\Delta r$
(controlling the discretization of the curves) used to estimate the
seismicity rate.
The inset shows the characteristic size of the cluster of foreshocks, measured
by the median of the distance between all foreshock-mainshock pairs,
as a function of the time before the mainshock. The solid line is a fit by a
power-law $R\sim t^H$ with $H=0.3$. Due to the large space-time window $T=10$
yrs and $R=500$ km used to select foreshocks, a large proportion of
the seismicity are background events, which induces a spurious migration
of seismicity toward the mainshock (see section \ref{secmigrobs}).}
\end{center}
\end{figure}

Another remarkable observation is presented in Figure \ref{forncR}
which shows the rate of foreshocks of type II
for mainshock magnitudes between $4$ and $4.5$, for different values of the
distance $R$ used to select aftershocks and foreshocks.
The inverse Omori law is observed up to $R \approx 200$ km,
and the duration of the foreshock sequences increases as $R$ decreases
due to the decrease of the effect of the background seismicity. Restricting
to the shortest distances $R$ to minimize the impact of
background seismicity, the inverse Omori laws can be observed up
to $10$ yrs before mainshocks, for foreshocks of type II (Figure \ref{Prfor}).
Thus, foreshocks  are not immediate precursors of mainshocks but
result from physical mechanisms of earthquake triggering
acting over very long times and large distances.

An important question concerns the relative weight of coincidental shocks,
i.e., early aftershocks triggered by a previous large earthquake, which
appear as foreshocks of type II to subsequent aftershocks (seen
as mainshocks of these foreshocks of type II). Such coincidental
shocks can give rise to an apparent inverse Omori law \cite{Shaw} when
averaging over all possible positions of ``mainshocks'' in the sequence,
without any direct interaction between these mainshocks
and preceding events viewed as their foreshocks.
Actually, these coincidental shocks form a minority of the total
set, because the fraction of shocks directly triggered by a mainshock
decays to negligible values beyond a few days for the range
of parameters of the ETAS that realistically fit the SCEC catalog
\cite{Felzer}.

\begin{figure}
\begin{center}
\psfig{file=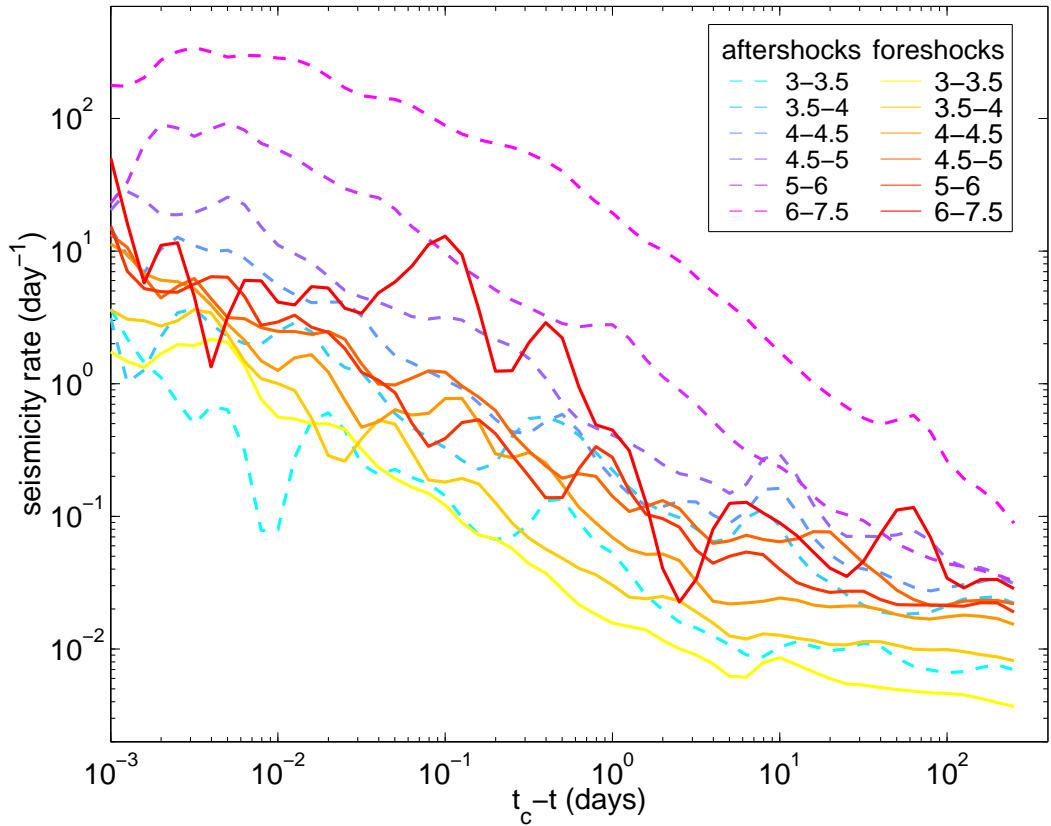,width=14cm}
\caption{ \label{SCECAFCR50T} Same as Figure \ref{SCECAFNCR50T} for
foreshocks of type I (defined in section \ref{techfore})
which have been selected using a space-time window
$R=R_2=50$ km and $T=1$ yr. The presented data and
the statistics for aftershocks are the same as in Figure \ref{SCECAFNCR50T}.
The total number of foreshocks of type I ranges from 1050 to
5462 depending on the mainshock magnitude. The same mainshocks are
used for the selection of aftershocks and of type I foreshocks.}
\end{center}
\end{figure}

Figure \ref{SCECAFCR50T} shows the rate of foreshocks of type I as a
function of $t_c-t$ and of aftershocks as a function of $t-t_c$, where
$t_c$ is the time of the mainshocks. The total number of foreshocks of
type I is much smaller that the number
of type II foreshocks for small mainshocks because a significant
fraction of foreshocks of type II are ``aftershocks'' of large $M>6$
earthquakes according to the usual definition and are therefore rejected
from the analysis of foreshocks of type I, which are constrained to be
smaller than their mainshock. There are much larger fluctuations
for foreshocks of type I than for foreshocks of type II due to the smaller
number of the former.
Nevertheless, type I and type II foreshocks clearly follow the same
trajectory of increased activity before a mainshock, suggesting that
type I foreshocks, like type II foreshocks, are triggers of the mainshock.

The exponent $p'$ of the inverse Omori law for foreshocks of type I is
approximately equal to the exponent of foreshocks of type II and to the
exponent $p$ of the direct Omori law for aftershocks.
The rate of foreshocks of type I increases slowly with the mainshock magnitude
but this increase is not due to a larger predictability of larger earthquakes,
as expected for example in the critical point theory \cite{Sorsammis}
and as observed in a numerical model of seismicity \cite{HSSS98}.
The increase of the number of type I foreshocks with the mainshock magnitude
can be reproduced faithfully in synthetic catalogs generated with the
ETAS model and is nothing but the consequence of the algorithmic rules
used to define foreshocks of type I.
Namely, by definition, bigger mainshocks are allowed to have bigger type I
foreshocks and therefore they have more type I foreshocks than smaller
mainshocks.
In other words, there is no physics but only statistics in the weak
increase of foreshocks of type I with the mainshock magnitude. Confirming
this concept, the inverse Omori law for foreshocks of type I becomes
independent of the mainshock magnitudes $M$ for large $M$, for which the
selection constraint has only a weak effect.
The dependence of the inverse Omori law for foreshocks of type I as a
function of distance $R$ used to select foreshocks is very similar (not shown)
to that shown for foreshocks of type II in Figure \ref{forncR}.
However, the duration of foreshock sequences is shorter for type I foreshocks
because the number of type I foreshocks is smaller than the number
of type II foreshocks due to the rules of foreshock selection.

\subsection{Modification of the magnitude distribution before a mainshock}

We have shown in \cite{Helmfore} that selecting and stacking
foreshock sequences in the ETAS model leads to a modification
of the magnitude distribution compared to the theoretical distribution.
We refer to \cite{Helmfore} for the derivation of the results
stated below, which is too involved to be reported here. The distribution
$P(m)$ of foreshock magnitudes is predicted to get an additive power law
contribution $q(t) dP(m)$ with an exponent $b'$ smaller than $b$ and with
an amplitude $q(t)$ growing as a power law of the time to the mainshock:
\be
P(m)=(1-q(t)) P_0(m)+ q(t)~ dP(m)~,
\label{Pmfor}
\ee
where $P_0(m)$ is the standard GR distribution $P_0(m) \sim 10^{-bm}$
and $dP(m) \sim 10^{-b'm}$ with $b'=b-\alpha$.
The amplitude $q(t)$ of the additive distribution $dP(m)$ in (\ref{Pmfor})
should increase as a power-law of the time to the mainshock according to
\be
q(t) \sim 1/(t_c-t)^{\theta { b' \over \alpha}}~.
\label{q}
\ee
This analytical prediction has been checked with extensive numerical
simulations of the ETAS model.
This change of the magnitude distribution for foreshocks does not
mean that foreshocks belong to a different population, but simply results
from the definition of foreshocks, which are only defined
as foreshocks after the mainshock occurred. The mechanism
giving rise to the change of distribution for foreshocks is explained
in \cite{Helmfore}.
Intuitively, the modification of the magnitude distribution of foreshocks
results from the increase of the number of triggered events with the
mainshock magnitude. There are few large earthquakes, but they trigger
many more earthquakes than smaller earthquakes. As a consequence,
a large fraction of mainshocks are triggered, directly or indirectly,
by large foreshocks. The proportion of large foreshocks is thus larger
than the proportion of large earthquakes in the whole population, and
gives an apparent lower $b$-value for foreshocks than for other earthquakes.
The magnitude distribution of triggering events is given by the product
$\rho(m) P_0(m) \sim 10^{\alpha m} 10^{-b m} \sim dP(m)$, which gives the
distribution of foreshock magnitudes for large $m$.

\begin{figure}
\begin{center}
\psfig{file=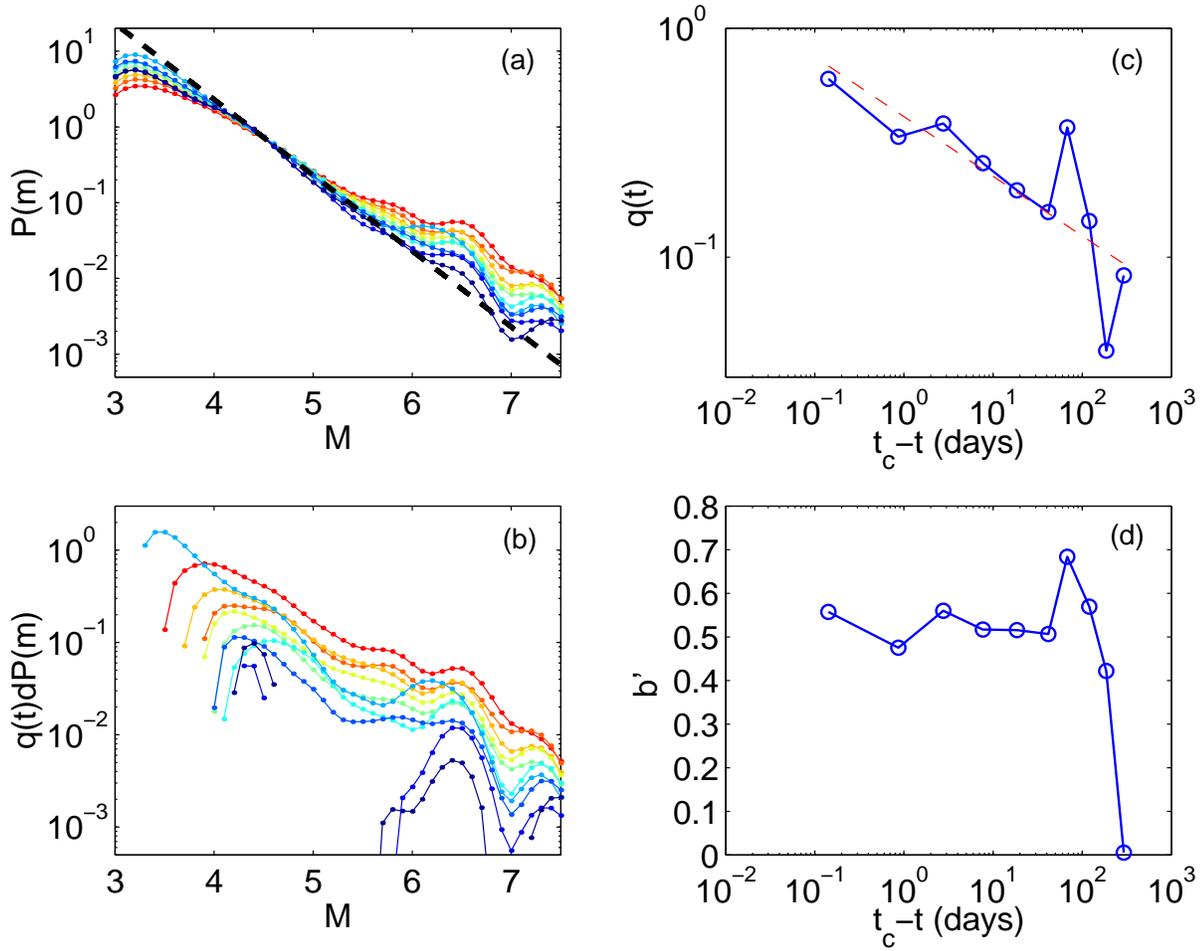,width=16cm}
\caption{ \label{PmSCEC2} (a) Magnitude distribution $P(m)$
of foreshocks of type II using a space window $R=20$ km,
for different time windows before the mainshock, ranging
  from black to gray as the time $t_c-t$ from the mainshock
decreases from 1 yr to 0.01 day.
Note  the progressive increase in the proportion of large earthquakes
by comparison to the normal distribution (dashed line, $b=1$).
Each curve contains the same number of events.
(b) difference $q(t)~dP(m)$ between the foreshock magnitude.
The foreshock magnitude
distribution is well fitted in the magnitude range $4\leq m \leq 7$
by the sum of two power-laws (\ref{Pmfor}), with an exponent
$b'\approx 0.5$ independently of the time from the mainshock.
The amplitude
$q(t)$ of the perturbation is shown in panel (c) and
the exponent $b'$ of $dP(m)$ is shown in panel (d). $q(t)$ is fitted
with a power law fit
as a function of $t_c-t$ with exponent $0.3 \pm 0.2$ shown as the
dashed line in panel (c). }
\end{center}
\end{figure}

We now test this prediction using the SCEC catalog on foreshocks
of type II of $M>3$ mainshocks, selected using $R=20$ km
and $T=1$ yr. The magnitude distribution $P(m)$ sampled at
different times before mainshocks is shown in
panel (a) of Figure \ref{PmSCEC2}, where the black to gray curves correspond
to times preceding mainshocks decreasing from $1$ year to $0.01$ day
with a logarithmic binning. As time approaches that of the mainshocks,
one can clearly observe that the tails depart more and more from the
standard GR power law $P_0(m)$ with $b=1.0 \pm 0.1$
estimated using the whole catalog and shown as the dashed line.
The perturbation $q(t) dP(m)$ in (\ref{Pmfor}) of the foreshock magnitude
distribution, shown in panel b) of Figure \ref{PmSCEC2},
can be estimated by fitting the prediction (\ref{Pmfor})
to the observed magnitude distribution of foreshocks, by inverting the
parameters $b'$ and $q(t)$ in  (\ref{Pmfor}) for different times before
mainshocks.
The obtained additive GR laws $q(t) dP(m)$  are compatible with pure
power laws with an approximately constant exponent
$b' =0.6 \pm 0.1$ shown in panel (d) of Figure \ref{PmSCEC2}, except at
very long times before the mainshock where it drops to 0 when the amplitude
$q(t)$ of the perturbation becomes too small. The amplitude $q(t)$
is shown in panel (c) and is compatible with a power law (\ref{q}) with
a fitted exponent $0.3 \pm 0.2$. These observations are in good
qualitative agreement with the predictions (\ref{Pmfor}) and (\ref{q})
on the nature of the modification of the GR law for foreshocks in terms
of a pure additive power law perturbation with an amplitude growing as a
power law of the time to the mainshocks.
  Quantitatively, $b'$ is marginally outside the $2\sigma$-confidence
interval for the prediction $b'=b-\alpha=0.2 \pm 0.2$ using the estimation
$\alpha=0.8 \pm 0.1$ given by [{\it Helmstetter}, 2003]. We attribute this
discrepancy to the dual impact of the incompleteness of the
catalog for small magnitudes after a large earthquake and to the smallness
of the statistics. We stress that the prediction (\ref{Pmfor}) with
$b'=b-\alpha$ has been verified with good precision in
synthetic catalogs which do not have these limitations
[{\it Helmstetter et al.}, 2003].
Using the best fitted value $b'=0.6$, we obtain a reasonable agreement
for the predicted exponent $\theta { b'/\alpha}$ and the fitted value
$0.3 \pm 0.2$ for the power law behavior of $q(t)$ using $\theta$ in
the range $0.2-0.4$.

Note that expression (\ref{Pmfor}) contains as a special case the model
in which the modification of the GR law occurs
solely by a progressive decrease of the $b$-value
as the time of the mainshock is approached (by putting $q(t)=1$
and allowing $b'$ to adjust itself as a function of time), as proposed
in \cite{Berg68,KK78,Molchan90,Molchan99}. Our quantitative analysis
clearly excludes this possibility while being completely consistent
with the mechanism embodied by the concept of triggered seismicity.
Although the foreshock magnitude distribution is not
a pure power-law but rather the sum of two power laws,
our results rationalize the reported decrease of $b$-value
before mainshocks \cite{Berg68,KK78,Molchan90,Molchan99}.
Indeed, with a limited number of events, the sum of two power
laws predicted by (\ref{Pmfor}) with an increasing weight of the
additive law $dP(m)$ as the time of the mainshock is approached
will be seen as a decreasing $b$-value when fitted with a single
GR power law.

\subsection{Migration of foreshocks}
\label{secmigrobs}

The last prediction discussed here resulting from the Hypothesis
is that foreshocks should migrate slowly toward the mainshock.
Note that the specification (\ref{first}) of the ETAS model
defined in section \ref{ETAS}
predicts no diffusion or migration if seismicity results
solely from direct triggering (first generation from mother to daughter).
Technically, this results from the separability of the space and time
dependence of $\phi_{M_i}(t-t_i, \vec r-\vec r_i)$.
In the ETAS model, diffusion and migration can be shown to result
from the cascade of secondary, tertiary (and so on) triggered seismicity,
akin to a (continuous-time) random walk with multiple steps
\cite{etasdif}, which
couples the space and time dependence of the resulting global seismicity rate.
This migration or anti-diffusion of the seismic activity
toward the mainshock is quantified by
the characteristic size $R$ of the cluster of foreshocks
which is predicted to decrease before the mainshock according to
\cite{etasdif,Helmfore}
\be
R \sim (t_c-t)^{H}~,
\label{Rfor}
\ee
with $H=\theta /\mu$ for $\mu<2$ where $\theta$ and $\mu$ are defined
in section \ref{ETAS}.
It is natural that the (sub-)diffusive exponent $H$ combines
the exponent $\theta$ (respectively $\mu$) of the time- (resp. space-)
dependent local processes (\ref{psidef}) and (\ref{phidef}).
This law (\ref{Rfor}) describes the localization
of the seismicity as the mainshock approaches, which is also observed in real
seismicity \cite{KK78,VS81}.

We use a superposed epoch analysis and stack all sequences of foreshocks
of type II synchronized at the time of the mainshock and with a common
origin of space at the location of each mainshock.
The analysis of the California seismicity presented in inset of Figure
\ref{Prfor} shows clearly a migration of the seismicity toward the mainshock,
confirmed  by the significant diffusion exponent $H=0.3 \pm 0.1$.
This value is compatible with the estimates $\theta \cong 0.2$ and
$\mu \cong 1$. We obtain the same pattern for type I and type II foreshocks,
but the figures for type I foreshocks have more noise due to the smaller
number of events.
However, this migration is likely to be an artifact of the background
activity, which dominates the catalog at long times and distances from
the mainshocks. Indeed, the shift in time from the dominance of the
background activity at large times before the mainshock to that of
the foreshock activity clustered
around the mainshock at times just before it may be taken as
an apparent inverse diffusion of the seismicity rate when using
standard quantifiers of diffusion processes (see \cite{etasdif}
for a discussion of a similar effect for the apparent diffusion of
aftershocks).

\section{Discussions and conclusions}

By defining the foreshocks of type II and by comparing
them with standard foreshocks of type I, we have
revisited the phenomenology of earthquake foreshocks using the
point of view of triggered seismicity formulated in our Hypothesis.
We have found that the most salient properties of foreshock sequences
are explained solely by the mechanism of earthquake triggering. This
validates the Hypothesis.

An important result is that the precursory modification of the seismic
activity before a mainshock is independent of its magnitude, as expected
by the triggering model with a constant magnitude distribution.
Therefore, large earthquakes are not more predictable than smaller
earthquakes on the basis of the power-law acceleration of the seismicity
before a mainshock or by using the modification of the magnitude distribution.

All these results taken together stress the importance of the multiple
cascades of earthquake triggering in order to make sense of the complex
spatio-temporal seismicity. In particular, our results do not use any of the
specific physical mechanisms proposed earlier to account for some of
the observations analyzed here. For instance, the ETAS model is different
from the receding stress
shadow model \cite{Bowking,Sorsammis}, from the critical earthquake model
\cite{Bowmanetal,JS,Sorsammis} and from the pre-slip model [{\it Dodge
et al.}, 1996], which in addition each addresses only a specific part
of the seismic phenomenology.
Our demonstration and/or confirmations of (i) the increase of rate of
foreshocks before mainshocks (ii) at large distances and (iii) up to decades
before mainshocks,  (iv) a change of the Gutenberg-Richter law from a concave
to a convex shape for foreshocks, and (v) the migration of foreshocks
toward mainshocks are reminiscent of, if not identical to, the precursory
patterns documented in particular by the Russian \cite{KeiMal,KBstate}
and Japanese \cite{Mogires} schools, whose physical origin has remained
elusive an/or controversial. The present work suggests that triggered
seismicity is sufficient to explain them.
The concepts and techniques and their variations developed here could be
applied to a variety of problems, such as to determine the origin
of financial crashes \cite{SMM03,JS03}, of major biological extinctions,
of change of weather regimes and of the climate, and in tracing the
source of social upheaval and wars \cite{SH03}.

The cascade model described here is sufficient to explain the properties
of foreshocks in time, space and magnitude. There may be however
other properties of foreshocks not explained by this model, such has
an unusual waveform or focal mechanism, that may
provide ways to distinguish foreshocks from other earthquakes
in real time
[{\it Dodge et al.}, 1996; {\it Kilb and Gomberg}, 1999].
For instance, {\it Dodge et al.} [1996] found that some foreshock
sequences in California were inconsistent with static stress triggering
of the mainshock, and concluded that these foreshocks were more likely
a by-product of an aseismic nucleation process.
Several points may reconcile these observations with the cascade
model. First, there are other mechanisms  such as dynamical effects
not taken into account in this study that may explain earthquake triggering.
Indeed, a large fraction of
aftershocks are inconsistent with static triggering by the mainshock.
Second, some foreshocks may be missing in the catalog used
and may have triggered the mainshock. Recall that, for $\alpha<b$,
the mainshock is more likely to be triggered by a small, possibly
undetected, earthquake [{\it Helmstetter}, 2003].

\acknowledgments
We are grateful to E. Brodsky, K. Felzer, J.-R. Grasso, H. Houston,
Y.Y. Kagan, G. Ouillon, J. Vidale and an anonymous referee for
stimulating discussions and constructive remarks that helped
improve the manuscript.
We are also grateful for the earthquake catalog data made available
by the Southern California Earthquake Data Center.
This work was partially supported by NSF-EAR02-30429, by
the Southern California Earthquake Center (SCEC) and
by the James S. Mc Donnell Foundation 21st
century scientist award/studying complex system.

{}

\end{article}

\clearpage

\clearpage

\clearpage

\clearpage

\end{document}